\documentclass[twocolumn,preprintnumbers,amsmath,amssymb]{revtex4}
\usepackage{epsfig,graphicx,hhline}
\usepackage{dcolumn}% Align table columns on decimal point
\usepackage{bm}% bold math

\begin{document}

\title{Elastic properties of quaternary oxyarsenide LaOFeAs
 as basic phase for new 26-52K superconductors}

\author{I.R. Shein$^*$ and A.L. Ivanovskii}

\affiliation {Institute of Solid State Chemistry, Ural Branch of
the Russian Academy of Sciences, 620219, Ekaterinburg, Russia}

\begin{abstract}
The full-potential linearized augmented plane wave (FP-LAPW) method with the generalized gradient approximation (GGA) for the exchange-correlation potential has been applied to predict the elastic properties of quaternary oxyarsenide LaOFeAs - the basic phase for the newly discovered 26-52K superconductors. The optimized lattice parameters, independent elastic constants (C$_{ij}$), bulk modulus B, compressibility $\beta$, and shear modulus G are evaluated and discussed. For the first time the numerical estimates of the elastic parameters for polycrystalline LaOFeAs ceramics are performed. Our analysis shows that LaOFeAs belong to the anisotropic mechanically stable systems; a parameter limiting the mechanical stability of this material is the shear modulus. In addition, LaOFeAs is a soft material (B $\sim$ 98 GPa) with high compressibility ($\beta$ $\sim$ 0.0102 1/GPa); lays at the border of a brittle behavior and adopts a high ionic contribution in intra-atomic bonding.\\

$^*$ E-mail: shein@ihim.uran.ru
\end{abstract}

\maketitle
    The elastic properties are of great importance for the material science of superconductors; 
on the other hand, the elastic constants can be linked to such important physical parameters 
of superconductors, as the Debye temperature $\Theta$$_D$ and the electron-phonon coupling 
constant $\lambda$ \cite{Ming}. In the past years, the elastic properties of various superconducting 
materials have been studied extensively, and some correlations between the superconducting 
critical temperature T$_c$ and mechanical parameters have been discussed. So, Hirsh \cite{Hirsch} 
has suggested that high T$_c$’s are associated with low values of the bulk modules B, i.e.
high compressibility $\beta$. Really, for the many groups of superconductors with enhanced 
critical temperatures (such as YBCO, MgB$_2$, MgCNi$_3$, boride-carbides and carbide-halides 
of the rare earth metals, REM$_2$B$_2$C and RE$_2$C$_2$X$_2$ etc) their bulk modules does 
not exceed B $\leq$ 200 GPa ($\beta$ $\geq$ 0.005 1/GPa) \cite{Ming,Hirsch,Meenakshi,Jeanicke-Rossler,
Wang,Shein,Guclu}. On the other hand, the superconducting transition (up to T$_c$ $\sim$ 11K) was
found for such hard and incompressible material as boron - doped diamond \cite{Ekimov,Takano}.\\
      Except scientific interest, mechanical properties appear extremely important for technology and
 various advanced applications of superconducting materials \cite{Scanlan}.\\
       In February 2008 the new layered superconductor: electron doped quaternary oxyarsenide
 LaOFeAs (LaO$_{1-x}$F$_x$FeAs, x$\sim$0.05-0.12) with T$_c$ at about 26K was reported \cite{Kamihara},
and this surprising discovery has stimulated much activity for search of the new superconducting 
materials based on this phase. It was found that also hole-doped oxyarsenide LaOFeAs (La$_{1-x}$Sr$_x$FeAs,
 by partial substitution of La by Sr) adopts superconductivity with T$_c$ at about 25K \cite{Wen}. Moreover
by replacing of La by others rare earth elements it seems to be possible to achieve the further enhancing 
of T$_c$ (41K - for CeO$_{1-x}$F$_x$FeAs \cite{Chen}, 43K - for SmO$_{1-x}$F$_x$FeAs \cite{Chen1}, 50-52K -
 for PrO$_{1-x}$F$_x$FeAs and NdO$_{1-x}$F$_x$FeAs \cite{Ren,Ren1}).\\
     Alongside with the intense efforts in synthesis of new doped quaternary oxypnictides and examining
of superconductivity parameters, some of their physical properties have been investigated. So, today 
the interplay between superconductivity and spin fluctuations, band structure, Fermi surface, and phonon 
spectra for some quaternary oxypnictides are discussed at the theoretical level \cite{Mazin,Kuroki,Singh,Zhang,Cao,Eschrig,Chen2,Marsiglio}; 
the photoemission spectra of SmO$_{0.85}$F$_{0.15}$FeAs are obtained \cite{Ou} and pressure experiments for LaO$_{1-x}$F$_x$FeAs (x=0.11) are performed \cite{Lu}.\\
     In this Report we focus on the elastic properties of monocrystalline quaternary oxyarsenide LaOFeAs as 
the basic phase for the newly discovered 26-52K superconductors. In addition the numerical estimates of
 the elastic parameters of the polycrystalline LaOFeAs ceramics, which may be important for the future 
applications of new superconducting materials are performed for the first time. \\
      The oxyarsenide LaOFeAs has a tetragonal (ZrCuSiAs-type structure, space group P4/nmm\cite{Zimmer}) layered structure,
 where Fe atoms are arrayed on a square lattice in the form of edge-shared tetrahedra FeAs$_4$. Each (Fe-As)$^-$ layer
 is sandwiched between (La-O)$^+$ layers. \\
    Our calculations based on density functional theory (DFT) with generalized gradient approximation (GGA) in PBE 
form \cite{Perdew} for the exchange-correlation potential and were performed by means of the full potential method with mixed
basis APW+lo (LAPW) implemented in the WIEN2k suite of programs \cite{Blaha}. The calculations are performed with full
lattice optimizations (lattice parameters and atomic coordinates) ; the self-consistent calculations were considered to
be converged when the difference in the total energy of the crystal did not exceed 0.001 mRy as calculated at
 consecutive steps, and the nonmagnetic state was treated. Other details are described in Ref \cite{Shein}.\\
   Firstly, the equilibrium lattice constants (a and c) were evaluated - in reasonable agreement with available data, 
Table I.\\
 Secondly, the values of six independent elastic constants (C$_{ij}$; namely C$_{11}$, C$_{12}$, C$_{13}$, C$_{33}$,
C$_{44}$ and C$_{66}$) for LaOFeAs were estimated by calculating the stress tensors on applying different deformations 
given on the equilibrium lattice of tetragonal unit cell and determining the dependence of the resulting change in energy on 
the deformation, Table I. All these elastic constants are positive and satisfy the well-known Born$'$s criteria for mechanically
 stable tetragonal crystal: C$_{11}$ $>$ 0, C$_{33}$ $>$ 0, C$_{44}$ $>$ 0, C$_{66}$ $>$ 0, (C$_{11}$ - C$_{12}$) $>$ 0,
 (C$_{11}$ + C$_{33}$ - 2C$_{13}$) $>$ 0 and {2(C$_{11}$ + C$_{12}$) + C$_{33}$ + 4C$_{13}$}  $>$ 0. The calculated 
anisotropic factor A= 2C$_{44}$/(C$_{11}$-C$_{12}$) is 0.65, i.e. LaOFeAs phase is mechanically anisotropic, because a
 value of A = 1 represents completely elastic isotropy, while values smaller or greater than 1 measure the degree of elastic anisotropy.\\
      The calculated elastic constants allow us to obtain the macroscopic mechanical parameters of LaOFeAs, namely bulk modulus
 (B) and shear modulus (G) - in two main approximations: Voigt (V) \cite{Voigt} and Reuss (R) \cite{Reuss}, in the following forms:\\

\begin{table}
\begin{center}
\caption{Calculated lattice constants (a, c, in $\AA$), cell volume (Vo, in $\AA^3$)
 elastic constants (C$_{ij}$, in GPa), bulk modulus (B, in GPa), shear modulus (G, in GPa)
 and B/G ratio for tetragonal monocrystalline LaOFeAs as well as some elastic parameters 
for polycrystalline LaOFeAs ceramics as obtained in the Voigt-Reuss-Hill approximation: 
bulk modulus (B$_{VRH}$, in GPa), compressibility ($\beta$ $_{VRH}$, in GPa$^{-1}$), 
shear modulus (G$_{VRH}$, in GPa), Young’s modulus (Y$_{VRH}$, in GPa)
 and Poisson’s ratio ($\nu$).}
\begin{tabular}{|c|c|c|c|}
\hline
a & 4.033 (4.036 [11]; 4.037 [30])\footnote{ in parentheses – the available experimental (Ref.\cite{Kamihara}) and calculated (Ref.\cite{Xu}) data are given} & B$_V$(B$_R$)\footnote{ in Voigt (and Reuss) approximations}& 98.5 (97.2)\\
\hline
c & 8.684  (8.739 [11]; 8.629 [30])$^a$& G$_V$(G$_R$)$^b$& 56.5 (55.9)\\
\hline
V$_0$& 141.2 & B$_V$/G$_V$ & 1.74\\
\hline
C$_{11}$ & 191.9 &B$_{VRH}$ &97.9\\
\hline
C$_{12}$ & 55.9 & $\beta$ $_{VRH}$ & 0.01022\\
\hline
C$_{13}$ & 61.6&  G$_{VRH}$& 56.2\\
\hline
C$_{33}$ & 144.8& Y$_{VRH}$ & 141.5\\
\hline
C$_{44}$ & 44.1&$\nu$ &0.259\\
\hline
C$_{66}$ & 77.9& &\\
\hline
\end{tabular}
\end{center}
\end{table}

\begin{equation*}
 B_{V}=\frac{1}{9}\lbrace{2(C_{11}+C_{12})+C_{33}+4C_{13}}\rbrace
\end{equation*}
\begin{equation*}
 G_{V}=\frac{1}{30}\lbrace{M+3C_{11}-3C_{12}+12C_{44}+6C_{66}}\rbrace
\end{equation*}
\begin{equation*}
 B_{R}=\frac{C^{2}}{M}
\end{equation*}
\begin{equation*}
 G_{R}=15\lbrace{\frac{18B_{V}}{C^{2}}+\frac{6}{C_{11}-C_{12}}+\frac{6}{C_{44}}+\frac{3}{C_{66}}}\rbrace^{-1}
\end{equation*}
\\
where C$^{2}$ = (C$_{11}$ + C$_{12}$)C$_{33}$ - 2C$_{13}$ and M = C$_{11}$ + C$_{12}$ + 2C$_{33}$ - 4C$_{13}$. 
The results obtained are summarized in Table I.\\
     Next, as LaOFeAs samples are usually prepared and investigated as polycrystalline ceramics [11,12,25,26] in the form of
aggregated mixtures of microcrystallites with a random orientation, it is useful to estimate the corresponding parameters 
for the polycrystalline material from the elastic constants of the single crystal. To this aim we utilize the Voigt-Reuss-Hill
(VRH) approximation. In this approach, according to Hill[33], the Voigt and Reuss averages are limits and the actual effective 
modules for polycrystals could be approximated by the arithmetic mean of these two bounds. Then, one can calculate the averaged 
compressibility ($\beta$ $_{VRH}$ = 1/B$_{VRH}$), Young modulus (Y$_{VRH}$) and from the B$_{VRH}$, G$_{VRH}$ and Y$_{VRH}$
one can evaluate the Poisson’s ratio ($\nu$). All these parameters are also listed in Table I. Certainly all of these estimations are
performed in limit of zero porosity of LaOFeAs ceramics.\\
      From our results we see that for LaOFeAs B$_{VRH}$ $>$ G$_{VRH}$; this implies that a parameter limiting the mechanical
stability of this material is the shear modulus G$_{VRH}$. \\
    The bulk modulus for LaOFeAs is rather small (at about 98 GPa), and is less than, for example, the bulk modules for other
known superconducting species: MgB$_2$ ($\sim$ 122-161 GPa [34]), MgCNi$_3$ ($\sim$172-210 GPa [35]), YBCO ($\sim$ 200 GPa), and YNi$_2$B$_2$C ($\sim$200 GPa [3]). Thus, compared with other superconductors, LaOFeAs is the soft material.\\
     According to criterion [36], the material is brittle if the B/G ratio is less than 1.75. In our case, for LaOFeAs B/G $\sim$ 1.74,
this suggests that this material lays at the border of a brittle behavior.\\
    Finally, the values of the Poisson ratio ($\nu$) for covalent materials are small ($\nu$ = 0.1), whereas for ionic materials a typical
value of $\nu$ is 0.25 [37]. In our case the value of $\nu$ for LaOFeAs is at about 0.26, i.e. a higher ionic contribution in intra-atomic
bonding for LaOFeAs phase should be assumed. Besides, for covalent and ionic materials, the typical relations between bulk and
shear modulus are G $\simeq$ 1.1B and G$\simeq$ 0.6B, respectively. For our case the calculated value of G$_V$/B$_V$ is 0.58,
also indicating that the ionic bonding is suitable for LaOFeAs.\\
   In summary, we have performed FLAPW-GGA calculations to obtain the elastic properties for mono- and polycrystalline oxyarsenide 
LaOFeAs as a basic phase for the newly discovered 26-52K superconductors. Our analysis shows that LaOFeAs belong to the anisotropic mechanically stable systems; a parameter limiting the mechanical stability of this material is the shear modulus. In addition, LaOFeAs is a
soft material (B$\sim$98 GPa) with high compressibility ($\beta$ $\sim$ 0.0102 1/GPa); lays at the border of a brittle behavior and adopts
a high ionic contribution in intra-atomic bonding.\\

%%%%%%%%%%%%%%%%%%%%%%%%%%%% Bibliography %%%%%%%%%%%%%%%%%%%%%%%%%%%

\end{document}